\documentclass[12pt,a4paper]{article}  
\usepackage{amsmath, amssymb, graphicx, amsfonts, latexsym, bbold, mathbbol}
\usepackage{slashed}
\usepackage[]{hyperref}
\usepackage{color}
\usepackage{scalerel}

\newcommand{\be}{\begin{equation}}
\newcommand{\ee}{\end{equation}}
\newcommand{\ba}{\begin{equation} \begin{aligned}}
\newcommand{\ea}{\end{aligned} \end{equation}}

\begin{document}

\begin{center}
{\LARGE \bf  Gauge-invariant coefficients in perturbative quantum gravity}
\vskip 1.2cm

Fiorenzo Bastianelli$^{\,a,b}$, Roberto Bonezzi$^{\,c}$, Marco Melis$^{\,a}$ 
\vskip 1cm

$^a${\em Dipartimento di Fisica e Astronomia ``Augusto Righi", Universit{\`a} di Bologna,\\
via Irnerio 46, I-40126 Bologna, Italy}\\[2mm]

$^b${\em   INFN, Sezione di Bologna, via Irnerio 46, I-40126 Bologna, Italy}\\[2mm]

$^c${\em Institute for Physics, Humboldt University Berlin,\\
 Zum Gro\ss en Windkanal 6, D-12489 Berlin, Germany}
 
\end{center}
\vskip .8cm

\abstract{Heat kernel methods are useful for studying properties of quantum gravity.
We recompute here the first three heat kernel coefficients in perturbative quantum gravity
with cosmological constant to ascertain which ones are correctly reported in the literature.
They correspond to the counterterms needed to renormalize the one-loop effective action in four dimensions. 
They may be evaluated at arbitrary dimensions $D$, in which case they identify only  
a subset of the divergences appearing in the effective action for $D\geq 6$. 
Generically, these coefficients depend on the gauge-fixing choice adopted in quantizing the Einstein-Hilbert action.  
However, they become gauge-invariant once evaluated on-shell, i.e. using 
Einstein's field equations with cosmological constant. We identify them and use them 
as a benchmark for checking alternative approaches to perturbative quantum gravity. 
One such approach describes the graviton in first-quantization through the use of the action 
of the  ${\cal N}=4$ spinning particle, characterized by four supersymmetries 
on the worldline and a set of worldline gauge invariances.
This description has been used for computing the gauge-invariant coefficients as well.
We verify their correctness at  $D=4$, but find a mismatch at arbitrary $D$ when 
comparing with the benchmark fixed earlier. We interpret this result as signaling that 
the path integral quantization of the ${\cal N}=4$ spinning particle should be amended.
We perform this task by fixing the correct counterterm that must be used in the worldline path integral
quantization of the ${\cal N}=4$ spinning particle to make it consistent in arbitrary dimensions.}

\section{\large Introduction}
A well-known approach to studying perturbative quantum gravity is to apply 
covariant quantization schemes and the background field method 
to the Einstein-Hilbert action, and then use heat kernel techniques for calculations, 
as pioneered by DeWitt \cite{DeWitt:1984sjp, DeWitt:2003pm}.
An alternative approach is to treat the graviton in first quantization, as in string theory.  
A model recently proposed for the graviton makes use of the ${\cal N}=4$ spinning particle 
and related BRST structure \cite{Bonezzi:2018box}.
It has been used in a path integral approach in \cite{Bastianelli:2019xhi} to study certain gauge-invariant  coefficients 
corresponding  to the on-shell, one-loop divergences of the effective action
of quantum gravity with cosmological constant.
These coefficients have been checked to be correct in $D=4$ dimensions.
Here, we wish to study the general case of arbitrary $D$. 

The most straightforward way to obtain these coefficients is to use the heat kernels of the differential operators 
appearing in the gauge-fixed Einstein-Hilbert action,
and compute the total heat kernel coefficients of quantum gravity out of them. 

We are interested in the first three coefficients, which we denote by $a_0$, $a_1$, $a_2$.
They are associated with the divergences of the one-loop effective action and should be renormalized away 
(additional divergences are present for $D\geq 6$).
The first one is universal and measures the number of physical degrees of freedom of the graviton. As for the
remaining ones, we have found different, inequivalent expressions in the literature. 
Most likely the different versions signal misprints. In any case, 
to ascertain which ones are correctly reported, we have decided to recompute them, 
verifying that the ones given in \cite{DeWitt:1984sjp}  and \cite{DeWitt:2003pm} 
which also differ between themselves are incorrect, 
while finding agreement with the ones computed more recently in \cite{Avramidi:2015pqa}.

These coefficients computed for a generic background metric depend on the gauge-fixing procedure adopted. 
They are expected to become gauge-invariant once restricted to Einstein spaces.
This is particularly evident for $a_0$, which is independent of the background metric and measures directly 
the number of physical polarizations of the graviton. 
We calculate these gauge-invariant coefficients explicitly, to have at hand a benchmark 
that any alternative attempt to perturbative quantum gravity should be able to reproduce.

With these coefficients at hand, we compare them with the ones obtained in \cite{Bastianelli:2019xhi} through
the ${\cal N}=4$ spinning particle and discover that there is agreement at $D=4$, 
while a mismatch appears for $D\neq 4$.
We take this result as suggesting that the path integral quantization of the 
${\cal N}=4$ particle should be reconsidered and improved. 
We do this by modifying the counterterm used in the path integral quantization.
The counterterm used in \cite{Bastianelli:2019xhi} was selected by taking into account the algebra of the four 
supersymmetries on the worldline that
seemed to indicate a specific form of the hamiltonian constraint, and it was found that the chosen counterterm could reproduce correctly
the results at $D=4$.
However, the founding principle of the ${\cal N}=4$ spinning particle description of the graviton is the BRST symmetry, 
which requires a specific coupling to the curvature to achieve nilpotency
and differs from the one used in \cite{Bastianelli:2019xhi}.
Modifying appropriately the counterterm, and precisely in the way dictated by the BRST symmetry,
we find that the gauge-invariant coefficients 
are correctly reproduced by the path integral of the ${\cal N}=4$ spinning particle at arbitrary $D$.

In this paper, we proceed as follows. In section \ref{sec2},
we describe the calculation of the heat kernel coefficients in perturbative quantum gravity at one loop, 
keeping the background metric arbitrary. In section \ref{sec3}, we identify the corresponding gauge-invariant coefficients by using Einstein's equations with cosmological constant.
In section \ref{sec4}, we reconsider the path integral for the ${\cal N}=4$ spinning particle and fix the correct counterterm that must go along with the
path integral in worldline dimensional regularization. 
This way the correct gauge-invariant coefficients are reproduced at arbitrary $D$
by the path integral on the circle of the ${\cal N}=4$ spinning particle.
We present our conclusions and outlook in section \ref{sec5}, leaving 
appendix \ref{A} for collecting useful formulae on one-loop effective actions and heat kernels.

\section{\large Heat kernel coefficients in perturbative quantum gravity}
\label{sec2}

Let us consider the Einstein-Hilbert action with cosmological constant $\Lambda$ 
for the metric $G_{\mu\nu}$ in $D$ euclidean dimensions 
\begin{equation}
    S[G_{\mu\nu}]=-\frac{1}{\kappa^2} \int d^Dx \sqrt{G} \,\Big[R(G)-2\Lambda \Big] 
\end{equation}
where $\kappa^2=16 \pi G_N$ is the gravitational coupling constant. 
The effective action can be studied using the background field method. One splits the metric as 
\begin{equation}
    G_{\mu\nu}(x)=g_{\mu\nu}(x)+h_{\mu\nu}(x)
\end{equation}
where $g_{\mu\nu}$ is an arbitrary background metric and $h_{\mu\nu}$ the quantum fluctuations.
To obtain the one-loop effective action, it is enough to expand the action 
at quadratic order in $h_{\mu\nu}$ 
\begin{equation}
    S[g+h]=\frac{1}{\kappa^2} \Big [S_0+S_1+S_2+\sum_{n=3}^\infty S_n \Big]
\end{equation}
where one finds
\begin{align}
\begin{split} 
    S_0 &=-\int d^Dx \sqrt{g} \big [R-2\Lambda\big ]\;, \\
    S_1 &=\int d^Dx \sqrt{g} \bigg [ h^{\mu\nu}\biggl(R_{\mu\nu}-\frac{1}{2}g_{\mu\nu}R
    +g_{\mu\nu}\Lambda\biggr) \bigg ]\;, \\
    S_2 &= \int d^Dx \sqrt{g} \bigg [ 
   -\frac{1}{4}h^{\mu\nu}(\nabla^2+2\Lambda - R)h_{\mu\nu}
   +\frac{1}{8}h(\nabla^2+2\Lambda-R)h
   -\frac{1}{2}\biggl(\nabla^\nu h_{\nu\mu} -\frac{1}{2}\partial_\mu h \biggr)^2 \\
        &\hskip2.35cm 
        -\frac{1}{2}\big(h^{\mu\lambda}h^\nu_\lambda-hh^{\mu\nu}\big )R_{\mu\nu}
              -\frac{1}{2}h^{\mu\lambda}h^{\nu\rho}R_{\mu\nu\lambda\rho}
                \bigg ]\;,
    \end{split}
\end{align}
with $h \equiv g^{\mu\nu}h_{\mu\nu}$.
Indices are raised and lowered with the background metric $g_{\mu\nu}$
and all curvature tensors and covariant derivatives are constructed using  $g_{\mu\nu}$. 
   
The gauge symmetries acting on $h_{\mu\nu}$  leave the background metric $g_{\mu\nu}$ invariant 
and must be gauge-fixed. It is useful to employ BRST methods and maintain the 
background gauge symmetry in the gauge-fixing terms as well.
This can be done by choosing a
weighted gauge based on the de Donder gauge-fixing function 
$f_\mu = \nabla^\nu h_{\nu\mu}-\frac{1}{2}\partial_{\mu}h$, which is a tensor under the background gauge symmetry. 
It gives rise to a total gauge-fixed action  $S_{tot}$ for the graviton $h_{\mu\nu}$ and 
ghost fields $b_\mu, c_\mu$ that at quadratic order reads
\begin{equation}
     S_{2,tot}[h,c,b]  = S_h[h] + S_{gh}[b,c]
\end{equation}
where
\begin{align}
S_{h} &=\int d^Dx \sqrt{g} \biggl[ -\frac{1}{4}h^{\mu\nu}(\nabla^2 +2 \Lambda-R) h_{\mu\nu} 
+ \frac{1}{8}h (\nabla^2 +2 \Lambda-R) h
\nonumber \\
 &\hskip 2.3cm -\frac{1}{2}\big(h^{\mu\lambda}h^\nu_\lambda -h h^{\mu\nu}\big)R_{\mu\nu} 
-\frac{1}{2}h^{\mu\lambda}h^{\nu\rho}R_{\mu\nu\lambda\rho}
\biggr] \;,
\label{6}
\\
 S_{gh}&= \int d^Dx \sqrt{g}\, b_\mu (-\delta^\mu_\nu \nabla^2 - R^\mu_\nu)c^\nu \;.
 \label{7}
\end{align}
The gauge-fixing procedure just described is standard, see for example \cite{DeWitt:1984sjp, DeWitt:2003pm}.
Here we have followed \cite{Bastianelli:2013tsa}, where additional details 
on the BRST procedure may be found. 

From these quadratic actions one can extract the invertible kinetic operators 
that lead to the one-loop effective action through the heat kernel. 
To this end, one can introduce a one-parameter family of metrics on the space of symmetric tensors $h_{\mu\nu}$:
\begin{equation}\label{general gamma}
\gamma_{(k)}^{\lambda\rho \mu\nu}=\frac12 \Big(g^{\lambda \mu}g^{\rho\nu}+g^{\lambda\nu}g^{\rho \mu}-k\,g^{\lambda\rho}g^{\mu\nu}\Big)\;,    
\end{equation}
which allow to write a norm for the graviton fluctuations as  
\be
||h||_{(k)}^2 =  \frac14
\int d^Dx \sqrt{g}
\,  h_{\lambda\rho }  \gamma_{(k)}^{\lambda\rho  \mu\nu}  h_{\mu\nu}   \;.
\label{9}
\ee
The metric \eqref{general gamma} has inverse
\begin{equation}
\gamma_{(k)\mu\nu\sigma\tau}=\frac12 \Big(g_{\mu\sigma}g_{\nu\tau}+g_{\nu\sigma}g_{\mu\tau}-\frac{2k}{kD-2}g_{\mu\nu}g_{\sigma\tau}\Big)\;,    
\end{equation}
which shows that $k=\frac{2}{D}$ is not allowed, as expected, since $\gamma_{\scaleobj{0.7}{\frac{2}{D}}}^{\lambda\rho\mu\nu}$ is the projector on the traceless subspace.
From now on we will choose $k=1$ following \cite{DeWitt:1984sjp} and work with
\be
 \gamma^{\lambda\rho \mu\nu}  
=\frac12 \Big(g^{\lambda \mu}g^{\rho\nu}+g^{\lambda\nu}g^{\rho \mu}-g^{\lambda\rho}g^{\mu\nu}\Big)\;,\quad \gamma_{\mu\nu\sigma\tau}=\frac12 \Big(g_{\mu\sigma}g_{\nu\tau}+g_{\nu\sigma}g_{\mu\tau}-\frac{2}{D-2}g_{\mu\nu}g_{\sigma\tau}\Big)
\label{8}\;.
\ee
This choice has the advantage of making the derivative part of the kinetic operator proportional to the identity in the tensor indices. This is also reflected in the fact that in flat space $\frac{\gamma_{\mu\nu\sigma\tau}}{p^2}$ is the graviton propagator in the Feynman-de Donder gauge.

Thus, we rewrite  the quadratic actions \eqref{6} and \eqref{7}
in the form
\begin{align}
S_{h} &=\int d^Dx \sqrt{g}
\,  \frac14 \, h_{\lambda\rho } \, \gamma^{\lambda\rho  \mu\nu} \, F_{\mu\nu}{}^{\sigma\tau} \, h_{\sigma\tau}
\label{10} 
\\
 S_{gh}&= \int d^Dx \sqrt{g}\, b_\mu \,
 \mathfrak{F}^\mu{}_\nu \,
 c^\nu 
\end{align}
to identify the differential operators 
\begin{align}
     F_{\mu\nu}{}^{\sigma\tau} &=-\frac12(\delta_{\mu}{}^\sigma \delta_{\nu}{}^\tau+\delta_{\mu}{}^\tau \delta_{\nu}{}^\sigma)(\nabla^2+2\Lambda-R) 
       \nonumber           \\
&\hskip .5cm
          - \frac{1}{D-2}g_{\mu\nu} g^{\sigma\tau} R +\frac{2}{D-2}g_{\mu\nu} R^{\sigma\tau} + g^{\sigma\tau}R_{\mu\nu} 
          \nonumber  \\
&\hskip .5cm
-\frac{1}{2}(\delta_\mu{}^\sigma R_\nu{}^\tau + \delta_\mu{}^\tau R_\nu{}^\sigma + \delta_\nu{}^\sigma R_\mu{}^\tau + \delta_\nu{}^\tau R_\mu{}^\sigma) 
- R_\mu{}^\sigma{}_\nu{}^\tau - R_\mu{}^\tau{}_\nu{}^\sigma \;, 
\\
\mathfrak{F}^\mu{}_\nu &=-( \delta^\mu{}_\nu\nabla^2+R^\mu{}_\nu)
\end{align}
whose determinants appear in the effective action for quantum gravity. 

Thus, let us consider the one-loop effective action. It is obtained by
computing the functional determinants of the above operators,
which we denote simply by $F$ and $\mathfrak{F}$. Using 
standard formulae, reviewed in appendix \ref{A}, we find for the one-loop effective action $\Gamma$
 \begin{align}
e^{-\Gamma}=  {\rm Det}^{-\frac12}  F  \ 
{\rm Det}\, \mathfrak{F}  
\end{align}
which in a proper-time representation leads to 
\begin{equation}
    \Gamma =- \frac12 \int_0^\infty \frac{dT}{T} \left (
    \text{Tr} \left [\text{e}^{-F T} \right]
    -2\text{Tr} \big [\text{e}^{-{\mathfrak{F}}T } \big] \right) 
\end{equation}
containing the heat kernels of the above operators.
Now, we insert the heat kernel expansions for small $T$, see eq. \eqref{63} of appendix \ref{A},
and identify the total heat kernel coefficients for quantum gravity
\begin{equation}
    \text{tr}[a_{n,tot}]=\text{tr} [a_n]  -  2\text{tr} [a_{n,gh}]
    \label{tot-a}
\end{equation}
where the first one is due to the graviton fluctuations and the second one to the ghosts.
This expansion is useful to identify the diverging terms of the effective action, 
but cannot be employed to obtain the finite terms
because of infrared divergences that appear 
from the upper limit of the proper time integration.

To evaluate the heat kernel coefficients, we specialize the general formulae of appendix \ref{A} to the operators $F$ and $\mathfrak{F}$, 
and find the following values. The graviton heat kernel coefficients from $F$ at arbitrary dimension $D$ are given by 
\begin{align}
    \begin{split}
        \text{tr}[a_{0}] &= \frac{1}{2}D(D+1)
        \label{a0D}
    \end{split}\\
    \begin{split}
        \text{tr}[a_{1}] &= -\frac{D(5D-7)}{12}R +D(D+1)\Lambda
    \end{split}\\
    \begin{split}
        \text{tr}[a_{2}] &= -\frac{D(2D-3)}{30}\nabla^2R + \frac{25D^3 - 145D^2 + 262D + 144}{144(D-2)}R^2 \\
    & -  \frac{D^3 - 181D^2 + 1438D - 720}{360(D-2)}R_{\mu\nu}R^{\mu\nu} + \frac{D^2 - 29D + 480}{360}R_{\mu\nu\sigma\tau}R^{\mu\nu\sigma\tau} \\
    & + D(D+1)\Lambda^2 - \frac{D(5D^2 - 17D + 14)}{6(D-2)}R\Lambda
    \end{split}
\end{align}
and at $D=4$ reduce to 
\begin{align}
    \begin{split}
        \text{tr}[a_{0}] \ &\xrightarrow{D=4} \ 10
        \end{split}\\
    \begin{split}
        \text{tr}[a_{1}] \ &\xrightarrow{D=4} \ -\frac{13}{3}R +20\Lambda
    \end{split}\\
    \begin{split}
        \text{tr}[a_{2}] \ &\xrightarrow{D=4} \ -\frac{2}{3}\nabla^2R + \frac{59}{36}R^2 
     -  \frac{55}{18}R_{\mu\nu}R^{\mu\nu} + \frac{19}{18}R_{\mu\nu\sigma\tau}R^{\mu\nu\sigma\tau} 
     + 20\Lambda^2 - \frac{26}{3}R\Lambda \;.
        \end{split}
\end{align}
One can recognize that eq. \eqref{a0D} counts the number of degrees of freedom of a symmetric rank-2 tensor 
with a non-vanishing trace. Now, one should add the contribution of the ghosts. 
However, as an aside, we notice that these coefficients are also useful as they stand. They describe 
the total coefficients of the system composed of gravity coupled to a complex spin 1 field 
and 4 real scalars: the determinant of the complex spin 1 field in the Feynman gauge compensates
that of the gravity ghosts, while the 4 scalars compensate the ghosts system for the complex spin 1 field.
As a simple check, one verifies that the total number of the physical degrees of freedom is
correctly reproduced by $\text{tr}[a_{0}]$.

Let us now turn to the coefficients due to the ghosts operator  $\mathfrak{F}$. They are given by
\begin{align}
    \begin{split} 
    \text{tr}[a_{0,gh}] &= D 
    \end{split}\\
    \begin{split}
        \text{tr}[a_{1,gh}] &= \frac{D+6}{6} R  
    \end{split}\\
    \begin{split}
        \text{tr}[a_{2,gh}] &= \frac{D+5}{30}\nabla^2R + \frac{D+12}{72}R^2
- \frac{D-90}{180} R_{\mu\nu}R^{\mu\nu} + \frac{D-15}{180}R_{\mu\nu\sigma\tau}R^{\mu\nu\sigma\tau}
    \end{split}
\end{align}
that in $D=4$ dimensions reduce to 
\begin{align}
    \begin{split}
    \text{tr}[a_{0,gh}] \ &\xrightarrow{D=4} \ 4
    \end{split}\\
    \begin{split}
        \text{tr}[a_{1,gh}]\  &\xrightarrow{D=4} \ \frac{5}{3}R
    \end{split}\\
    \begin{split}
        \text{tr}[a_{2,gh}] \ &\xrightarrow{D=4}\  \frac{3}{10}\nabla^2R + \frac{2}{9}R^2
+ \frac{43}{90} R_{\mu\nu}R^{\mu\nu} - \frac{11}{180}R_{\mu\nu\sigma\tau}R^{\mu\nu\sigma\tau} \;.
    \end{split}
\end{align}

Finally, the total coefficients for quantum gravity are obtained by evaluating \eqref{tot-a} and read
\begin{align}
\begin{split}
    \text{tr}[a_{0,tot}] &= \frac{D(D-3)}{2}
    \label{eq:31}
\end{split}\\
\begin{split}
    \text{tr}[a_{1,tot}] &= - \frac{5D^2-3D+24}{12}R + D(D+1)\Lambda
\end{split}\\
\begin{split}
    \text{tr}[a_{2,tot}] &= -\frac{2D^2-D+10}{30}\nabla^2R + \frac{25D^3-149D^2+222D+240}{144(D-2)}R^2 \\
    & - \frac{D^3-185D^2+1806D-1440}{360(D-2)}R_{\mu\nu}R^{\mu\nu} + \frac{D^2-33D+540}{360}R_{\mu\nu\sigma\tau}R^{\mu\nu\sigma\tau} \\
    & + D(D+1)\Lambda^2 - \frac{5D^3-17D^2+14D}{6(D-2)}R\Lambda
    \label{eq:33}
\end{split}
\end{align}
reducing at $D=4$ to 
\begin{align}
\begin{split}
    \text{tr}[a_{0,tot}] \ & \xrightarrow{D=4} \ 2
\end{split}\\
\begin{split} 
    \text{tr}[a_{1,tot}] \ & \xrightarrow{D=4} \ -\frac{23}{3}R + 20\Lambda
\end{split}\\
\begin{split} 
    \text{tr}[a_{2,tot}] \ & \xrightarrow{D=4} \ -\frac{19}{15}\nabla^2R + \frac{43}{36}R^2 - \frac{361}{90}R_{\mu\nu}R^{\mu\nu} + \frac{53}{45}R_{\mu\nu\sigma\tau}R^{\mu\nu\sigma\tau} + 20\Lambda^2 - \frac{26}{3}R\Lambda \;.
\end{split}
\end{align}

Einstein's field equations have not been used at any step in our calculations and
the results are valid for any background. However, they depend 
on the choice of the gauge-fixing terms, i.e. they are not BRST invariant.

Let us compare them with the ones that are found in the literature.
One may check that, at arbitrary dimension $D$ and vanishing cosmological constant, 
some of them differ from the ones reported in eqs. (16.80)-(16.82) of \cite{DeWitt:1984sjp} 
and in eq. (35.168) of the second volume of \cite{DeWitt:2003pm}, which also differ between themselves. 
Nevertheless, they are identical to the ones recently computed in \cite{Avramidi:2015pqa}, 
which we thus consider as the correct ones.

We should stress that these terms depend in general on the gauge-fixing procedure adopted, which however
is the same in the references cited above. 
To obtain truly gauge-independent coefficients one should evaluate them on-shell, i.e. using a 
background that satisfies Einstein's equations. We do this in the next section.

\section{\large Gauge-invariant coefficients in perturbative quantum gravity}
\label{sec3}

To derive gauge independent coefficients, we use background metrics 
that correspond to Einstein spaces, that by definition 
satisfy  Einstein's equations with cosmological constant $\Lambda$
\begin{equation}
    R_{\mu\nu}-\frac{1}{2}g_{\mu\nu} R + g_{\mu\nu} \Lambda =0
\end{equation}
allowing to 
relate the cosmological constant  and the Ricci tensor to the Ricci scalar 
\begin{equation} \label{eqn:relations}
\Lambda=\frac{D-2}{2D}R\;,  \ \ \ \ R_{\mu\nu}=\frac{1}{D}\, g_{\mu\nu} R\;.
\end{equation}
Using these relations, we find that the coefficients  $a_{n,tot}$ in eqs. \eqref{eq:31}--\eqref{eq:33} reduce to 
\begin{align}
\begin{split}
\label{39}
    \text{tr}[a_{0,tot}] &= \frac{D(D-3)}{2}
\end{split}\\
\label{40}
\begin{split}
    \text{tr}[a_{1,tot}] &= \frac{1}{12}(D^2-3D -36) R
 \end{split}\\
\label{41}
\begin{split}
    \text{tr}[a_{2,tot}] &= \frac{(D+5)(5D^2-42D-144)}{720D}R^2 + \frac{D^2-33D+540}{360}R_{\mu\nu\sigma\tau}R^{\mu\nu\sigma\tau}
\end{split}
\end{align}
with values at $D=4$ given by 
\begin{align}
\label{eqn:42}
\begin{split}
    \text{tr}[a_{0,tot}] & \ \xrightarrow{D=4} \ 2
\end{split}\\
\label{eqn:43}
\begin{split}
    \text{tr}[a_{1,tot}] & \ \xrightarrow{D=4} \ -\frac{8}{3}R
\end{split}\\
\label{eqn:44}
\begin{split}
    \text{tr}[a_{2,tot}] & \ \xrightarrow{D=4} \ -\frac{29}{40}R^2 + \frac{53}{45}R_{\mu\nu\sigma\tau}R^{\mu\nu\sigma\tau}
    \;.
\end{split}
\end{align}

As these terms are evaluated on-shell, they should not depend
on the gauge chosen. They identify gauge-invariant coefficients. 
They sit on the divergences of the effective action and must be renormalized away.
Let us briefly comment on this point, reviewing some old literature.
In $D=4$ and setting the cosmological constant to zero, also the Ricci scalar vanishes, and from \eqref{eqn:43} one finds 
that $\text{tr}[a_{2,tot}]\sim R_{\mu\nu\sigma\tau}^2$, which becomes a total derivative proportional to the Euler density on Einstein spaces. 
 Being a total derivative, it can be dropped from the effective action.
This makes the logarithmic divergences due to \eqref{eqn:44}
absent from the one-loop effective action, thus reproducing the famous result  of 
t' Hooft and Veltman \cite{tHooft:1974toh}, according to which quantum gravity is finite at one-loop 
(more precisely, it is free of logarithmic divergences, as the quartic divergence from $a_{0,tot}$
contributes to the cosmological constant;
in any case, quantum gravity remains renormalizable at one loop).
This result does not hold anymore at two loops and pure quantum gravity becomes non-renormalizable,
as shown by Goroff and Sagnotti \cite{Goroff:1985th} and checked by van de Ven
\cite{vandeVen:1991gw}. 
Adding a cosmological constant, the logarithmic divergence at $D=4$ 
does not vanish on-shell anymore, even dropping the total derivative corresponding to the Euler density.
The precise coefficient from  \eqref{eqn:44}
coincides with the one obtained long ago by Christensen and Duff \cite{Christensen:1979iy}.
Similarly, the term $\text{tr}[a_{1,tot}]$ in $D=4$  coincides with the one calculated in 
\cite{Bastianelli:2013tsa}.

At arbitrary $D$, and with $D\geq 6$, the above gauge-invariant coefficients form only a subset 
of the possible divergences of perturbative quantum gravity. As they are gauge-invariant, 
any formulation of quantum gravity should be able to reproduce them,
independently of the scheme chosen in the calculation.  

An alternative formulation treats the graviton in first quantization. 
A mechanical action useful for describing the graviton
is the $\mathcal{N}=4$ spinning particle. When supplemented with a set of gauge symmetries,
the $\mathcal{N}=4$ spinning particle has only the graviton in the physical spectrum at arbitrary $D$.
It has been used in \cite{Bastianelli:2019xhi} to compute the gauge-invariant coefficients discussed above.
 A direct comparison shows that the results are correctly reproduced in 4 dimensions, 
but differ at arbitrary $D$ in the terms proportional to $R$ and $R^2$.
We take this fact as suggesting that the path integral quantization of the 
$\mathcal{N}=4$ spinning particle developed in 
\cite{Bastianelli:2019xhi} is correct only at $D=4$, but needs an improvement for arbitrary $D$.
We discuss this issue next.

\section{\large Worldline path integral for the graviton}  
\label{sec4}

In flat space, a relativistic particle with $\cal N$-extended local supersymmetry
on the worldline describes a particle of spin $s=\frac{\cal N}{2}$ in four dimensions,
as suggested in \cite{Berezin:1976eg} and demonstrated explicitely in 
\cite{Gershun:1979fb, Howe:1988ft}.
One obtains spin 2, the graviton, by setting ${\cal N}=4$,
but couplings to nontrivial backgrounds
have been proven difficult to achieve for sufficiently large $\cal N$,
including the case of the graviton,
see refs. \cite{Kuzenko:1995mg, Bastianelli:2008nm}
for some attempts. The worldline path integral performed on the circle in \cite{Bastianelli:2007pv}
could just reproduce the physical degrees of freedom of the graviton.
At that time it was also realized that the gauging of the full $R$-symmetry 
group of the $\cal N$-extended supersymmetry,
the group $SO({\cal N})$ that rotates the 
${\cal N}$ real supercharges, could have been relaxed to a subgroup
without destroying the unitarity of the model, a fact used in
\cite{Pashnev:1990cf} and \cite{Bastianelli:2015tha}
to describe multiplets of particles of different spins.

The understanding of allowed couplings to background fields was
improved by using BRST methods.
As discussed in \cite{Dai:2008bh}, BRST techniques
give a way of introducing the Yang-Mills couplings for the   
${\cal N}=2$ spinning particle, which describes the propagation of a particle of spin~1.
Following similar strategies, it was found  in \cite{Bonezzi:2018box} how the graviton can be coupled 
to a curved background that satisfies Einstein's field equations 
with or without cosmological constant, see also \cite{Bonezzi:2020jjq} for further extensions.

The same BRST construction was used in \cite{Bastianelli:2019xhi}
to define a worldline path integral for the graviton by using 
the ${\cal N}=4$ spinning particle.  The path integral takes into account the gauging of worldline translations,
supersymmetries, and a parabolic subgroup of the  $SO({\cal N})$ $R$-symmetry group.
Gauging of translations and supersymmetries guarantees the unitarity of the model, 
while the gauging of a maximal parabolic subgroup of $SO({\cal N})$
leaves only irreducible spin 2 states in the physical Hilbert space. 
The model includes two additional Chern-Simons couplings fixed in such a way
to describe the graviton in arbitrary dimensions $D$.
Performing the path integral on a circle, one finds a worldline representation of the 
one-loop effective action of the graviton  $\Gamma [g_{\mu\nu}] $ with the schematic form
\be
\Gamma[g_{\mu\nu}] =  \int_{S^1}
\frac{\mathcal{D}G\,\mathcal{D}X^\mu}{\mathrm{Vol(Gauge)}}\, {\rm e}^{-S[X^\mu,G;g_{\mu\nu}]} \;.
\ee
where the particle action $S[X^\mu,G;g_{\mu\nu}]$ depends on the worldline gauge fields
$G=(e, \chi_i, a_{ij})$, and on the coordinates $X^\mu=(x^\mu,\psi^\mu_i)$ that contain
the true worldline coordinates $x^\mu$ plus their supersymmetric partners $\psi^\mu_i$
with $i=1,..,4$, while $g_{\mu\nu}$ 
is the background metric. The BRST symmetry underlying the model makes sure that the path integral 
is correct when the background metric $g_{\mu\nu}$ is on-shell, meaning that it must correspond 
to a metric of an Einstein space. Upon gauge fixing, 
and using complex combinations of the real fermions, that we still denote by $\psi^\mu_i$ but
with a redefined internal index $i$ taking only two values ($i=1,2$),
the path integral takes the following concrete form
\begin{equation}\label{3.4}
\Gamma[g_{\mu\nu}] = -\frac12 \int_0^\infty\frac{dT}{T}
\int_0^{2\pi}\frac{d\theta}{2\pi} \int_0^{2\pi} \frac{d\phi}{2\pi} \; \mu(\theta, \phi)
\int_P {\cal D}x \int_A D\bar \psi D\psi\  {\rm e}^{-S_{g}}
\ee
where the gauge-fixed, nonlinear sigma model action for the graviton is
\be
S_{g}=\int_0^1 d\tau\Big[\,\frac1{4T} \,g_{\mu\nu}\dot x^\mu \dot x^\nu +\bar\psi^{ia}(
	\delta_i^j D_\tau - \hat a_i^j)\psi_{ja}
		-T R_{abcd}\,\bar \psi^a\cdot \psi^b\, \bar \psi^c\cdot \psi^d +2T\,V_0\Big] 
\label{47}	\ee
	where
\be \hat a_i^j={\left(\begin{array}{cc}
		\theta &0\\
		0&\phi
	\end{array}\right)}
	\ee
contains the two moduli $\theta$  and $\phi$. We have used flat indices 
on the worldline complex fermions $\psi^a_i$, so that $D_\tau$ is the covariant derivative with the spin connection, 
while a dot denotes contraction on the internal indices.
The scalar potential $V_0\sim R$ is an order $\hbar^2$ improvement term to be discussed shortly.
The measure  $\mu(\theta, \phi)$ on the moduli space $(\theta, \phi)$ coming from
the Faddeev--Popov determinants is given by
\be
\mu(\theta, \phi) =\frac12 \left(2\cos \frac{\theta}{2}\right)^{-2} \left(2\cos \frac{\phi}{2}\right)^{-2}\, 
2i\,\sin\tfrac{\theta+\phi}{2}\left(2i\,\sin\tfrac{\theta-\phi}{2}\right)^2
\,
{\rm e}^{-iq(\theta+\phi)}
\ee
and includes the Chern-Simons coupling  $q=\frac{3-D}{2}$,
needed to select the graviton for arbitrary spacetime dimensions $D$.
We refer to  \cite{Bastianelli:2019xhi} for 
further details on the particle action and its gauge fixing on the circle.

In \cite{Bastianelli:2019xhi}  the scalar potential $V_0$ in eq. \eqref{47} was taken to be given by $V_0=-\tfrac{D+2}{8(D-1)}R$
upon consideration of the ${\cal N}=4$ supersymmetry algebra. 
In addition, the path integral for the nonlinear sigma model must be regularized, 
see \cite{Bastianelli:2006rx} for a review on this issue. The counterterms needed 
in models with extended supersymmetries can be found in~\cite{Bastianelli:2011cc}. 
From the latter reference, one extracts the counterterm $ V_{\rm CT}=\frac18 R $ needed in 
the worldline dimensional regularization for the ${\cal N}=4$ model.
Once added to $V_0$, it gives a total potential
\begin{align}
	V=	V_{\rm CT}+ V_0=
	\left(\frac18 -\frac{D+2}{8(D-1)}\right)R = -\frac{3}{8(D-1)}R =: \omega R
	\label{50}
\end{align}
that must be used in \eqref{47} instead of $V_0$ when using worldline dimensional regularization.
This was done in \cite{Bastianelli:2019xhi}, where a
 perturbative calculation for small $T$ delivered the divergences of the effective action
 in the form
\begin{align}
\Gamma[g_{\mu\nu}] =
		-\frac12 \int_0^\infty\frac{dT}{T}
	\int d^Dx\, \frac{\sqrt g}{(4\pi T)^{\frac D2}}\,
	\Big\langle\!\!\Big\langle {\rm e}^{-S_{int}}\Big\rangle\!\!\Big\rangle
	\label{3.7}
\end{align}
where $\langle\langle \dots \rangle \rangle$ denotes the perturbative
corrections of the path integral and subsequent modular integration, giving for Einstein spaces the answer 
\begin{equation}
     \Big\langle\!\!\Big\langle e^{-S_{int}}\Big\rangle\!\!\Big\rangle =
     a_0 + a_1T + a_2 T^2 
     +O(T^3)
\end{equation}
with
\begin{align}
\begin{split}
a_{0} &= \frac{D(D-3)}{2}
\end{split}\\
\begin{split}
a_{1} &= \left [\frac{D^2}{24}(5-24 \omega)+ D \Big(3\omega -\frac{39}{24}\Big) \right ] R
 \end{split}\\
\begin{split}
a_{2} &=  \Bigg [\frac{D^2}{576} (5-24\omega)^2
- D \Big ( 3\omega^2 -\frac{13}{4} \omega +\frac{1583}{2880} \Big) + \frac{121}{120} +\frac{2}{D}   \Bigg ]  R^2 + \frac{D^2-33D+540}{360}R_{\mu\nu\sigma\tau}R^{\mu\nu\sigma\tau}\;.
\end{split}
\end{align}
From these results, one verifies that the value of $\omega$ defined by eq. \eqref{50}
reproduces only the coefficients at $D=4$ but not those at arbitrary $D$, just compare with eqs. \eqref{39}--\eqref{41}.

One may guess that the mismatch is due to an incorrect identification
of the potential $V$ in eq. \eqref{50} used
in \cite{Bastianelli:2019xhi}
for getting the above result,
as for the rest the construction of the path integral stands on solid principles.
This conjecture is correct.
One may fix $\omega$ by requiring that it reproduces the expected coefficient $a_1$ in eq. \eqref{40}, finding 
\be
\omega =\frac18 -\frac{1}{D} \;.
\ee
Then, one verifies that 
with this value of $\omega$ also $a_2$ comes out correctly.
This is a nontrivial check, as there is no left-over freedom in defining the path integral.

The correct value of $\omega$ could have been deduced also from first principles.
The BRST analysis of ref. \cite{Bonezzi:2018box}
requires a value of $V_0$ in the hamiltonian constraint to be given by
$V_0= -\frac{1}{D} R$ to achieve nilpotency of the BRST charge on the 
relevant physical subspace of the full BRST Hilbert space, see eq. (5.12)
of \cite{Bonezzi:2018box}. Adding the counterterm $ V_{\rm CT}=\frac18 R $ 
required by the worldline dimensional regularization of the ${\cal N}=4$ model
gives a total potential
\be
V=\left (\frac18 -\frac{1}{D} \right )R
\ee
with the correct value of $\omega$ used above.
This modification makes the worldline path integral for the graviton correct in any spacetime dimension.

It is perhaps surprising that both values of the potential $V$ coincide at $D=4$. This is probably due to the fact that the potential $V_0=-\frac{D+2}{8(D-1)}\,R$ was found in \cite{Bastianelli:2008nm} upon demanding closure of the supersymmetry algebra on maximally symmetric spaces. Since the model analyzed in \cite{Bastianelli:2008nm} does propagate a graviton in $D=4$ (but not in arbitrary dimensions), one may understand why the above value for $V_0$ coincides with the correct one $-\frac{1}{D}\,R$ in this case.  

\section{\large Conclusions} 
\label{sec5}

Heat kernel methods have proved to be useful 
to study properties of QFT, like one-loop effective actions, anomalies, dressed propagators,
etc.,
see for example the recent applications aimed at  
finding the correct trace anomalies of chiral fermions \cite{Bastianelli:2016nuf, Bastianelli:2019zrq}. 
We have used them again here to recompute the full set of divergences of the one-loop  
effective action of quantum gravity in $D=4$, but keeping the spacetime dimension arbitrary 
(for $D\geq 6$ there are additional divergences).
Evaluating them on-shell allowed us to identify gauge-invariant coefficients, 
that may be used as a benchmark for testing alternative formulations of quantum gravity.
We have used them straight away to check 
the path integral construction of the graviton in first-quantization, which employs
the ${\cal N}=4$ spinning particle. A mismatch 
of the coefficients at $D\neq 4$  has prompted 
us to improve on the path integral construction to make it consistent in arbitrary dimensions.

This first-quantized approach to gravity extends the scope of worldline methods \cite{Schubert:2001he} to include gravity as a quantum theory.   
Recent developments on worldline approaches to QFT have addressed the study of 
dressed propagators and more general properties in QED \cite{Ahmadiniaz:2017rrk, Ahmadiniaz:2020wlm, Ahmadiniaz:2021gsd, Feal:2022iyn}.
We hope that additional gravitational applications may extend the usefulness of the method further.
In that regard, it might be useful to consider 
the extension of quantum gravity to complex spaces, 
which may be studied by using the simpler $U({\cal N})$ particles considered 
in \cite{Marcus:1994mm, Bastianelli:2009vj,Bastianelli:2011pe, Bastianelli:2012nh}.
Finally, it might be useful to extend the calculation of the gauge-invariant coefficients 
to include $a_3$, generalizing the results of ref. \cite{VanNieuwenhuizen:1977ca} to arbitrary dimensions and with a 
nonvanishing cosmological constant.

\section*{Acknowledgments}

R.B. would like to thank the Department of Physics and Astronomy of the University of Bologna for kind hospitality during the initial stage of this work.
The work of R.B. is funded by the European Research Council (ERC) under the European Union’s
Horizon 2020 research and innovation programme (grant agreement No 771862).


\appendix 

 \section{Effective action and heat kernel} \label{A}
 We recapitulate here standard formulae for the one-loop effective action and heat kernel.
  
 The one-loop effective action is linked to the determinant of a differential operator $H$ that may
 depend on various background fields. For our purposes, it is enough to consider 
 the sufficiently general case of a real scalar field $\phi$
 of mass $m$ in curved space, coupled to a non-abelian gauge field and a Lie algebra valued scalar potential.
 The euclidean action is taken to be quadratic in the scalar field 
 and of the general form
 \be
 S[\phi] = \int d^D x \sqrt{g} \ \frac12 \phi^T (H + m^2) \phi
 \ee
 with $H$ the second order differential operator mentioned above.
 The corresponding  one-loop effective action $\Gamma$ depends on the background fields contained in $H$.
 It is obtained by a Gaussian path integral  
 \begin{align}
e^{-\Gamma}= \int D\phi \, e^{-S[\phi]} = {\rm Det}^{-\frac12} ( H +m^2 )
= e^{-\frac12 {\rm Tr} \ln ( H +m^2)}
\end{align}
which leads, using a proper-time representation and dropping a constant term, to 
\be
\Gamma= \frac12  {\rm Tr} \ln ( H +m^2)= 
-\frac12  \int_0^\infty \frac{dT}{T}\, e^{-m^2T} {\rm Tr}\, e^{- HT}
\label{ea}
\ee
where ${\rm Tr}\, e^{- HT}$ is the functional trace of the heat kernel of $H$.

For $H$ we consider an operator of the form
\be
H = -\nabla^2 + V
\ee
with $V$ a Lie algebra valued potential. The laplacian $\nabla^2=\nabla^\mu \nabla_\mu$ is 
constructed with the gravitational and gauge covariant derivative $\nabla_\mu =D_\mu + W_\mu$, where
$D_\mu$ is the covariant derivative containing the usual metric connection while
$W_\mu$ is the Lie algebra valued gauge field. The covariant derivative $\nabla_\mu$ satisfies
\be
[\nabla_\mu,\nabla_\nu]\phi = {\cal F}_{\mu\nu}\phi  \;, \qquad [\nabla_\mu,\nabla_\nu] A^\lambda 
= R_{\mu\nu}{}^\lambda{}_\rho A^\rho
\ee
where $\phi$ is the scalar field in a real representation of the gauge group 
and $A^\mu$ a vector field invariant under the gauge group.
From these commutators one reads off the definition of the various curvatures. 
In our conventions, $R_{\mu\nu}= R_{\lambda\mu}{}^\lambda{}_\nu$ and 
 $R= R^\mu_\mu>0$ on a sphere.

The trace of the heat kernel corresponding to $H$ is given in perturbation theory for small $T$
\be
{\rm Tr} \left [  e^{- H T} \right ]
= \int \frac{d^Dx \sqrt{g(x)}}{(4 \pi T)^\frac{D}{2}}\, {\rm tr}\, (a_0(x)+ a_1(x) T + a_2(x) T^2+\cdots)
\label{63}
\ee
where the symbol ``tr'' denotes the trace on the remaining discrete matrix indices 
corresponding to the chosen representation of the gauge group and
\ba
a_0(x) &= {\mathbb 1} 
\cr
a_1(x) &=    \frac16 R \hskip .4mm  {\mathbb 1}  - V \cr
a_2(x) &=  \frac{1}{180} (R^2_{\mu\nu\lambda\rho}-R_{\mu\nu}^2) {\mathbb 1} 
+ \frac{1}{72} \left ( R \hskip .4mm  {\mathbb 1}   - 6V \right)^2 
+\frac{1}{30} \nabla^2 \left ( R \hskip .4mm  {\mathbb 1}   - 5V \right) + \frac{1}{12} {\cal F}_{\mu\nu}^2 
\label{SdW}
\ea
are the heat kernel coefficients (also known as Seeley-DeWitt coefficients).
As $V$ is matrix valued, $\nabla_\mu V= \partial_\mu V+ [W_\mu, V]$.
 
After inserting \eqref{63} into \eqref{ea},
one finds the following derivative expansion of the effective action
\be
\Gamma= -\frac12  \int d^Dx \sqrt{g(x)}
\int_0^\infty \frac{dT}{T}\, 
\frac{e^{-m^2T}}{(4 \pi T)^\frac{D}{2}} 
\, 
{\rm tr}\, (a_0(x)+ a_1(x) T + a_2(x) T^2+\cdots) \;.
\ee
The mass term guarantees convergence at the upper limit of the proper-time integral.
Infrared divergences arise in the massless limit and invalidate 
in that case this derivative expansion.
Ultraviolet divergences arise instead from the lower limit of the proper-time integration, i.e. at  $T\to 0$. 
For example, in $D=4$ one may verify that $a_0$, $a_1$, and $a_2$
all lead to  UV divergences, namely quartic, quadratic, and logarithmic divergences, respectively.
These UV divergences must be renormalized away. As already mentioned, 
in the massless limit the above expansion is not valid for getting finite terms of the effective action,
but it is still useful to recognize the explicit form of the UV divergences, 
as in the application to quantum gravity described in the main text.
In particular, ${\rm tr}\, a_0$ counts the number of degrees of freedom, is normalized to 1 
for a single real scalar field, and contributes to the divergence of the cosmological constant.


\end{document}